\documentclass[conference]{IEEEtran}
\IEEEoverridecommandlockouts
\usepackage{cite}
\usepackage{amsmath,amssymb,amsfonts}
\usepackage{algorithmic}
\usepackage{graphicx}
\usepackage{textcomp}
\usepackage{xcolor}
\usepackage{bm}
\newcommand{\rom}[1]{%
	\textup{\uppercase\expandafter{\romannumeral#1}}%
}
\def\BibTeX{{\rm B\kern-.05em{\sc i\kern-.025em b}\kern-.08em
    T\kern-.1667em\lower.7ex\hbox{E}\kern-.125emX}}
\begin{document}

\title{Modelling power grids as pseudo adaptive networks\\
%{\footnotesize \textsuperscript{*}Note: Sub-titles are not captured in Xplore and should not be used}
\thanks{This work was supported by the German Research Foundation DFG, Project Nos. 411803875 and 440145547.}
}

\author{\IEEEauthorblockN{Rico Berner}
\IEEEauthorblockA{\textit{Institut f\"ur Theoretische Physik} \\
%\textit{Technische Universit\"at Berlin}\\
\textit{Institut f\"ur Mathematik} \\
\textit{Technische Universit\"at Berlin} \\
Berlin, Germany \\
ORCID 0000-0003-4821-3366}
\and
\IEEEauthorblockN{Serhiy Yanchuk}
\IEEEauthorblockA{\textit{Institut f\"ur Mathematik} \\
\textit{Technische Universit\"at Berlin} \\
Berlin, Germany \\
ORCID 0000-0003-1628-9569}
\and
\IEEEauthorblockN{Eckehard Sch\"oll}
\IEEEauthorblockA{\textit{Institut f\"ur Theoretische Physik} \\
\textit{Technische Universit\"at Berlin} \\
%\textit{Bernstein Center for Computational Neuroscience} \\
Berlin, Germany \\
\textit{Potsdam Institute for Climate Impact Research} \\
Potsdam, Germany \\
ORCID 0000-0002-7318-2672}
}
\maketitle

\begin{abstract}
Power grids, as well as neuronal networks with synaptic plasticity, describe real-world systems of tremendous importance for our daily life. The investigation of these seemingly unrelated types of dynamical networks has attracted increasing attention over the last decade. In this work, we exploit the recently established relation between these two types of networks to gain insights into the dynamical properties of multifrequency clusters in power grid networks. For this, we consider the model of Kuramoto-Sakaguchi phase oscillators with inertia and describe the emergence of multicluster states. Building on this, we provide a new perspective on solitary states in power grid networks by introducing the concept of pseudo coupling weights.
\end{abstract}


%--------------------------------------------------
% Introduction
%--------------------------------------------------
\section{Introduction}\label{sec:intro}
Complex networks describe various processes in nature and technology, ranging from physics and neuroscience to engineering and socioeconomic systems. Of particular interest are power systems as well as micro and macro power grids~\cite{SAU98a,FIL08a,SCH16o}. It was shown that simple low-dimensional models capture certain aspects of the short-time dynamics of power grids very well~\cite{DOE13,NIS15,AUE16}. In particular, the model of phase oscillators with inertia, also known as swing equation, has been widely used in works on synchronization of complex networks~\cite{DOE14,ROD16} and as a paradigm for the dynamics of modern power grids~\cite{ROH12,COR13,MOT13a,MEN14,WIT16,AUE17,MEH18,JAR18,SCH18c,SCH18i,TAH19,HEL20,KUE19,TOT20,MOL21}.

Over the last years, studies on models of oscillators with inertia have revealed a plethora of common dynamical scenarios with adaptive network models of coupled oscillators. These scenarios include solitary states~\cite{JAR18,TAH19,HEL20,BER20c}, frequency clusters~\cite{BEL16a,BER19,BER19a,TUM19}, chimera states~\cite{OLM15a,KAS17}, hysteretic behavior and non-smooth synchronization transitions~\cite{OLM14a,ZHA15a,BAR16a,TUM18}. Moreover, hybrid systems with phase dynamics combining inertia with adaptive coupling weights have been investigated, for instance, to account for a changing network topology due to line failures~\cite{YAN17a}, to include voltage dynamics~\cite{SCH14m} or to study the emergence of collective excitability and bursting~\cite{CIS20}.

Despite the apparent qualitative similarities of the two types of models, only recently their quantitative relationship has been discovered~\cite{BER20a}. In this paper, we show implications of the relation for dynamical power grid models and provide future perspectives for the research on power grid stability and control.

The paper is organized as follows. In Sec.~\ref{sec:model}, we introduce the Kuramoto-Sakaguchi model with inertia. In the subsequent section~\ref{sec:dynRel}, we briefly review the analytic relation between power grid models and adaptive networks and introduce the concept of the pseudo coupling matrix. In Sec.~\ref{sec:multiCl_KwI} we show the emergence of a multicluster for oscillators with inertia. Subsequently, in Sec.~\ref{sec:solitaryPG}, we show how the concept of pseudo coupling weights can be used to study solitary states in realistic power grid networks. Finally, in Sec.~\ref{sec:conclusions}, we summarize the results and give an outlook.

%--------------------------------------------------
% Phase oscillator models
%--------------------------------------------------
\section{Kuramoto-Sakaguchi model with inertia}\label{sec:model}
The model that is considered throughout this paper is given by $N$ coupled phase oscillators with inertia~\cite{JAR18}
\begin{align}\label{eq:KwI_2order}
	M\ddot{\phi}_i +\gamma\dot{\phi}_i & = P_i - \sigma \sum_{j=1}^N a_{ij}\sin(\phi_i-\phi_j+\alpha),
\end{align}
with phase $\phi_i(t)\in(0,2\pi]$ and phase velocity $\dot{\phi}_i(t)=\frac{\mathrm{d}\phi_i(t)}{\mathrm{d}t}$ of the $i$th node ($i=1,\dots,N$), corresponding to generators and loads~\cite{SCH14m}. The $\phi_i(t)$, $\dot{\phi}_i(t)$ are defined relative to a rotation with reference power line frequency $\omega_G$, e.g., $50$ Hz for European power grid. The parameter $M$ is the inertia coefficient, $\gamma$ is the damping constant, $\sigma$ is the overall coupling strength, and $P_i$ is the power injected or consumed at node $i$ (related to the natural frequency $\omega_i = {P_i}/{\gamma}$). The connectivity between the oscillators is described by the entries $a_{ij}\in\{0,1\}$ of the adjacency matrix $A$. Further, the coupling function $\sin(\phi+\alpha)$ is parameterized by the phase lag parameter $\alpha$~\cite{SAK86}. The phase lag can be interpreted as part of complex impedance~\cite{HEL20}.

%--------------------------------------------------
% Dynamical relation between the phase oscillator models
%--------------------------------------------------
\section{Relating power grid models to adaptive networks}\label{sec:dynRel}
In the following, we present a relation between phase oscillator models with inertia and systems with adaptive coupling weights, and provide an extension for higher order power grid models including voltage dynamics.
\subsection{Pseudo coupling weights: The link between inertia and network adaptivity}
Consider $N$ adaptively coupled phase oscillators~\cite{KAS17,BER19,BER20b}
\begin{align}
	\dot{\phi}_i &= \omega_i + \sum_{j=1}^N a_{ij}\kappa_{ij} f(\phi_i-\phi_j), \label{eq:APO_phi}\\
	\dot{\kappa}_{ij} & = -\epsilon\left(\kappa_{ij} + g(\phi_i - \phi_j)\right), \label{eq:APO_kappa}
\end{align}
where $\phi_{i}\in[0,2\pi)$ represents the phase of the $i$th oscillator ($i=1,\dots,N$), $\omega_i$ is its natural frequency, and $\kappa_{ij}$ is the coupling weight of the connection from node $j$ to $i$. Further, $f$ and $g$ are $2\pi$-periodic functions where $f$ is the coupling function and $g$ is the adaptation rule, and $\epsilon$ is the adaptation rate that is usually chosen to be small ($\epsilon \ll 1$). The entries $a_{ij}$ of the adjacency matrix $A$ describe again the connectivity of the network.

In order to find the relation between \eqref{eq:KwI_2order} and \eqref{eq:APO_phi}--\eqref{eq:APO_kappa}, we first write Eq.~\eqref{eq:KwI_2order} in the form
\begin{align}
	\dot{\phi}_i &= \omega_i + \psi_i, \label{eq:KwI_1order_phi}\\
	\dot{\psi}_i & = -\frac{\gamma}{M}\left({\psi}_i + \frac{\sigma}{\gamma}\sum_{j=1}^N a_{ij}\sin(\phi_i - \phi_j+\alpha)\right), \label{eq:KwI_1order_psi}
\end{align}
where $\psi_i$ is the deviation of the instantaneous phase velocity from the natural frequency $\omega_i$. We observe that this is a system of $N$ phase oscillators~\eqref{eq:KwI_1order_phi} augmented by the adaptation~\eqref{eq:KwI_1order_psi} of the frequency deviation $\psi_i$. Note that the coupling between the phase oscillators is realized in the frequency adaptation which is different from the classical Kuramoto system~\cite{KUR84}. As we know from the theory of adaptively coupled phase oscillators~\cite{KAS17,BER19}, a frequency adaptation can also be achieved indirectly by a proper adaptation of the coupling matrix. 

In order to introduce coupling weights into system~\eqref{eq:KwI_1order_phi}--\eqref{eq:KwI_1order_psi}, we express the frequency deviation $\psi_i$ as the sum $\psi_i = \sum_{j=1}^N a_{ij}\chi_{ij}$ of the dynamical power flows $\chi_{ij}$ from the nodes $j$ that are coupled with node $i$. The power flows are governed by the equation $\dot{\chi}_{ij}  = -\epsilon\left({\chi}_{ij} + g(\phi_i - \phi_j)\right)$, where $g(\phi_i - \phi_j)\equiv \sigma \sin(\phi_i - \phi_j+\alpha)/\gamma$ are their stationary values~\cite{SCH18i} and $\epsilon =\gamma/M$. It is straightforward to check that $\psi_i$, defined in such a way, satisfies the dynamical equation~\eqref{eq:KwI_1order_psi}.

As a result, we have shown that the swing equation~\eqref{eq:KwI_1order_phi}--\eqref{eq:KwI_1order_psi} can be  written as the following system of adaptively coupled phase oscillators
\begin{align}
	\dot{\phi}_i &= \omega_i+\sum_{j=1}^N a_{ij}\chi_{ij}, \label{eq:KwI_pseudo_phi}\\
	\dot{\chi}_{ij} & = -\epsilon\left({\chi}_{ij} + g(\phi_i - \phi_j)\right). \label{eq:KwI_pseudo_kappa}
\end{align}
The obtained system corresponds to~\eqref{eq:APO_phi}--\eqref{eq:APO_kappa} with coupling weights $\chi_{ij}$ and coupling function $f(\phi_i-\phi_j) \equiv 1$. The coupling weights form a pseudo coupling matrix $\chi$. Note that the base network topology $a_{ij}$ of the phase oscillator system with inertia Eq.~\eqref{eq:KwI_2order} is unaffected by the transformation.

With the introduction of the pseudo coupling weights $\chi_{ij}$, we embed the $2N$ dimensional system~\eqref{eq:KwI_1order_phi}--\eqref{eq:KwI_1order_psi} into a higher dimensional phase space. In~\cite{BER20a} it was shown that the dynamics of the higher dimensional system \eqref{eq:KwI_pseudo_phi}--\eqref{eq:KwI_pseudo_kappa} is completely governed by the system \eqref{eq:KwI_1order_phi}--\eqref{eq:KwI_1order_psi} on a $2N$ dimensional invariant submanifold, thereby establishing a mathematically rigorous relation.

Let us discuss the physical meaning of the coupling weights $\chi_{ij}$. For this, we consider the power flows $F_{ij}$ from node $j$ to node $i$ given by $F_{ij}=-g(\phi_i - \phi_j)$~\cite{SCH18i}. Then each $\chi_{ij}$ is driven by the power flow from $j$ to $i$. In particular, for constant $F_{ij}$, $\chi_{ij}\to F_{ij}$ asymptotically as $t\to \infty$ on the timescale $1/\epsilon$. Therefore, $\chi_{ij}$ acquires the meaning of a dynamical power flow.

The obtained result suggests that the power grid model is a specific realization of adaptive neuronal networks. Indeed, in the following, we proceed one step further and show that more complex models for synchronous machines can be represented as adaptive network as well.

\subsection{Swing equation with voltage dynamics as adaptive network with metaplasticty}\label{sec:swing}
Here we generalize the results of the previous subsection for the swing equation with voltage dynamics~\cite{SCH14m,TAH19}:
\begin{align}
	M\ddot{\phi}_i +\gamma\dot{\phi}_i & = P_i + \sum_{j=1}^N E_i E_j a_{ij}h(\phi_i-\phi_j),
	\label{eq:extended_voltagedyn}\\
	m_i \dot{E}_i & = - E_i + E_{f,i} + \sum_{j=1}^N a_{ij}E_j v(\phi_i - \phi_j),\label{eq:extended_voltagedyn_E}
\end{align}
where the additional dynamical variable $E_i$ is the voltage amplitude. The functions $h$ and $v$ are $2\pi$-periodic, and $m_i$ and $E_{f,i}$ are machine parameters~\cite{SCH14m,TAH19}. All other variables and parameters are as in~\eqref{eq:KwI_2order}. 

Equations \eqref{eq:extended_voltagedyn}--\eqref{eq:extended_voltagedyn_E}
can be rewritten as an adaptive network  \eqref{eq:KwI_pseudo_phi}--\eqref{eq:KwI_pseudo_kappa}
supplemented by Eq.~\eqref{eq:extended_voltagedyn_E} where $g(\phi)\equiv - E_i E_jh(\phi)/\gamma$ and  $\epsilon = \gamma/M$. 
%We note that the phase space of~\eqref{eq:extended_voltagedyn}--\eqref{eq:extended_voltagedyn_E} is $3N$ dimensional.
For this, in analogy to Sec.~\ref{sec:dynRel}A, we  write~\eqref{eq:extended_voltagedyn}--\eqref{eq:extended_voltagedyn_E} as
\begin{align}
	\dot{\phi}_i &= \omega_i+\sum_{j=1}^N a_{ij}\chi_{ij},\\
	\dot{\chi}_{ij} & = -\frac{1}{M_i}\left(\gamma{\chi}_{ij} - E_i E_jh(\phi_i - \phi_j)\right),\label{eq:extended_KwI_pseudoCoup}\\
	m_i \dot{E}_i & = - E_i + E_{f,i} + \sum_{j=1}^N a_{ij}E_j v(\phi_i - \phi_j),
\end{align}
where we introduce the coordinate changes $\chi_{ij}\to \chi_{ij}+ {P_i}/{\gamma}$, $E_i\to E_i+E_{f,i}$ and set $\omega_i = {P_i}/{\gamma}$. Due to the voltage dynamics~\eqref{eq:extended_voltagedyn_E}, the adaptation function $g(\phi)=E_i(t) E_j(t)h(\phi)$ in~\eqref{eq:extended_KwI_pseudoCoup} possesses additional adaptivity. This kind of meta-adaptivity (meta-plasticity) is of importance in neuronal networks~\cite{ABR96,ABR08a} as well as for neuromorphic devices~\cite{JOH18}.

%--------------------------------------------------
% Mixed frequency cluster states in phase oscillator models with inertia
%--------------------------------------------------
\section{Mixed frequency cluster states in phase oscillator models with inertia}\label{sec:multiCl_KwI}
In this section, we provide a novel viewpoint of the emergence of multifrequency cluster states for phase oscillator models with inertia. In such a state all oscillators split into $C$ groups (called clusters) each of which is characterized by a common cluster frequency $\Omega_\mu$. In particular, the temporal behavior of the $i$th oscillator of the $\mu$th cluster ($\mu=1,\dots,C$) is given by $\phi_i^\mu (t)= \Omega_\mu t + \rho^{\mu}_i + s^{\mu}_i(t)$ where $\rho^{\mu}_i\in[0,2\pi)$ and $s^{\mu}_i(t)$ are bounded functions describing different types of phase clusters characterized by the phase relation within each cluster~\cite{BER19}. Various types of multicluster states including the special subclass of solitary states have been extensively described for adaptively coupled phase oscillators~\cite{KAS17,BER19a,BER20c}.

%\begin{center}
\begin{figure}
	\centering
	\includegraphics[width=0.8\columnwidth]{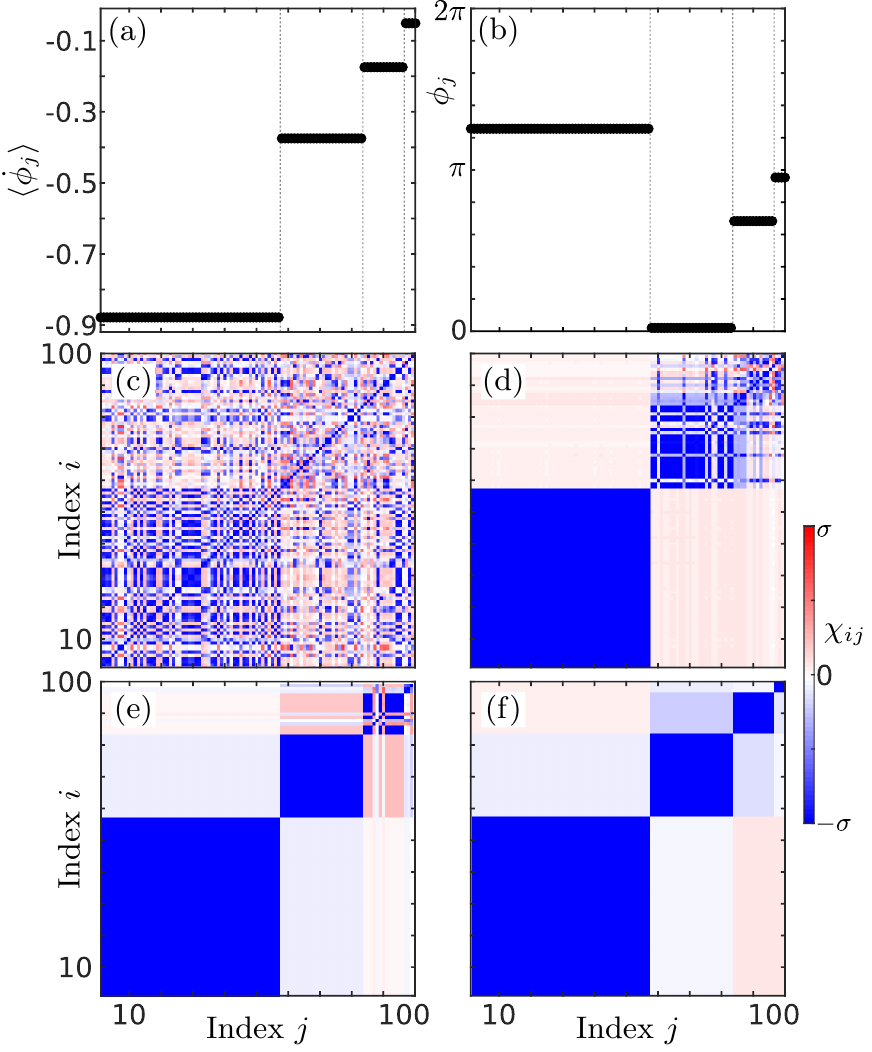}
	\caption{\label{fig:MC_KwI}
		Hierarchical multicluster state in a network of coupled phase oscillators with inertia. The panels (a,b) show the (temporally averaged) mean phase velocities $\langle\dot{\phi}_j\rangle$, and phase snapshots $\phi_j(t)$, respectively, at $t=10000$. All groups of oscillators characterized by a common mean phase velocity are separated by gray dotted lines. The temporal evolution of the pseudo coupling matrix $\chi_{ij}(t)$ is presented in (c) $t=100$, (d) $t=1750$, (e) $t=5000$, and (f) $t=10000$. Starting from an incoherent state in panel (c), the largest cluster is formed first (d), and the other clusters are then successively formed depending on their size (e),(f). In (a-f) the oscillator indices are sorted in increasing order of their mean phase velocity. The state is found by numerical integration of~\eqref{eq:KwI_2order} with identical oscillators $P_i=0$ and uniform random initial conditions $\phi_i(0)\in(0,2\pi)$, $\psi_i(0)\in(-0.5,0.5)$. Parameters: globally coupled network $a_{ij}=1$ for all $i\ne j$, $M=1$, $\gamma=0.05$, $\sigma=0.016$, $\alpha=0.46\pi$, $N=100$.}
\end{figure}
%\end{center}

In Figure~\ref{fig:MC_KwI}, we present a $4$-cluster state of in-phase synchronous clusters on a globally coupled network. As we know from the findings for adaptive networks~\cite{BER19}, (hierarchical) multicluster states are built out of single cluster states whose frequency scales approximately with the number $N_\mu$ of elements in the cluster. In the zeroth-order expansion in $\gamma$, the collective cluster frequencies are given by $\Omega_\mu\approx-\sigma N_\mu\sin \alpha$. Multicluster states exist in the asymptotic limit ($\gamma\to 0$) also for networks of phase oscillators with inertia if the cluster frequencies are sufficiently different meaning the clusters are hierarchical in size. Remarkably, the pseudo coupling matrix displayed in Fig.~\ref{fig:MC_KwI}(f) shows the characteristic block diagonal shape that is known for adaptive networks. In particular, the oscillators within each cluster are more strongly connected than the oscillators between different clusters.

Another observation for multicluster states in networks of phase oscillators with inertia is their hierarchical emergence. As reported in~\cite{KAS17} for adaptive networks, the clusters emerge in a temporal sequence from the largest to the smallest. In Fig.~\ref{fig:MC_KwI}(c-f), we show that this particular feature is also found in phase oscillators with inertia. 

%We have shown that the findings for multicluster states for adaptively coupled phase oscillators can be transferred to networks of phase oscillators with inertia. Moreover, we find that the adaptive feature given by a the inertia term is sufficient to obtain stable frequency clustering. Note that the system considered above is homogeneous. However, heterogeneous real-world networks can be treated with the methods from adaptive networks as well. To show this, in the next section, we analyze the dynamical characteristics of solitary states in the German ultra-high voltage power grid network~\cite{EGE16}.

%--------------------------------------------------
% Solitary states in the German ultra-high voltage power grid network
%--------------------------------------------------
\section{Solitary states in the German ultra-high voltage power grid}\label{sec:solitaryPG}
In this section, we show that multifrequency cluster states, as discussed in Fig.~\ref{fig:MC_KwI}, may also occur in real power grid networks, which are heterogeneous in contrast to the identical oscillators treated in the previous section. For the simulation, we consider the Kuramoto model with inertia given by Eq.~\eqref{eq:KwI_2order}. The network structure and the power distribution are taken from the ELMOD-DE data set provided in~\cite{EGE16}.

\begin{figure}
	\centering
	\includegraphics[width=0.8\columnwidth]{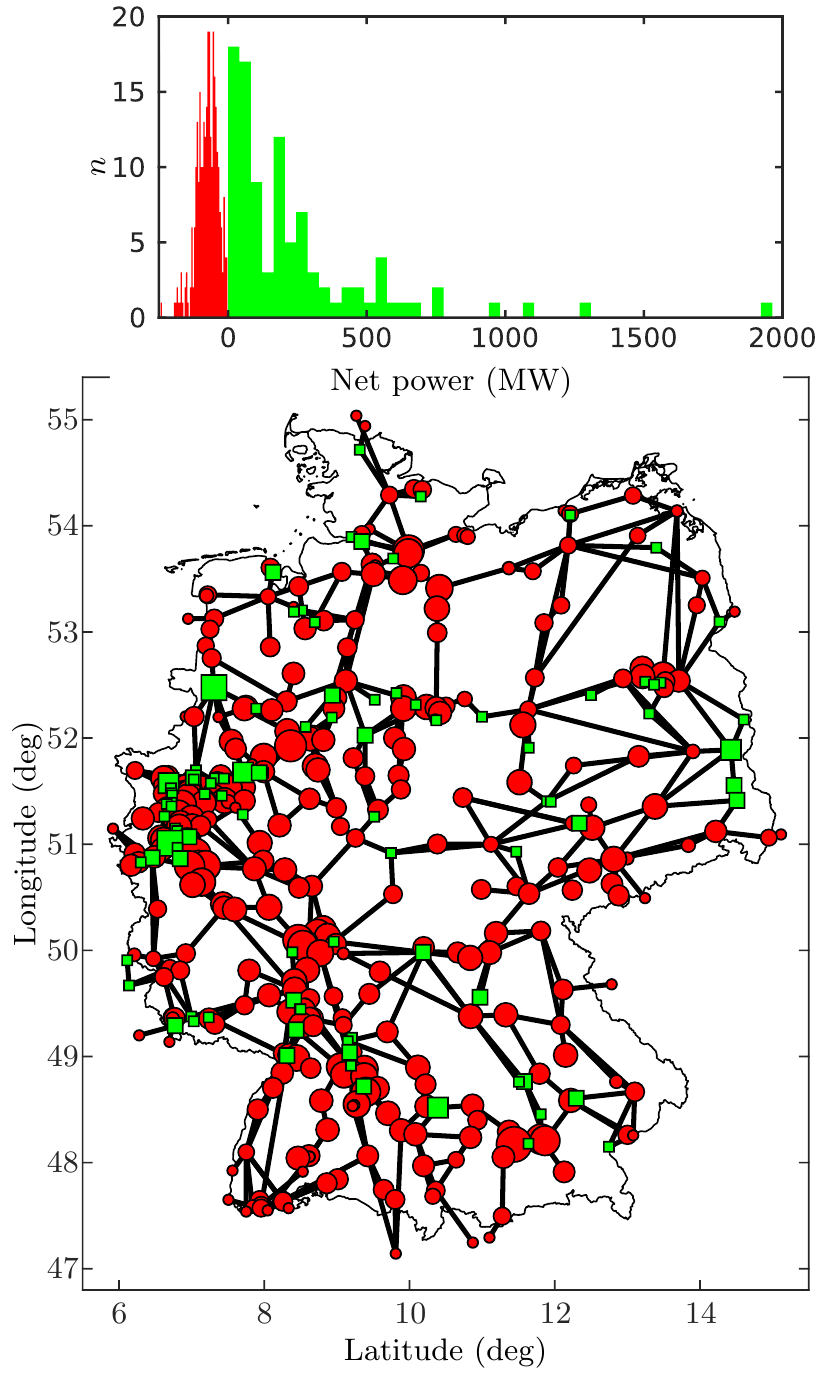}
	\caption{\label{fig:powerGridGermany} Map of the German ultra-high voltage power grid consisting of $95$ net generators (green squares) and $343$ net consumers (red circles) connected by $662$ bidirectional transmission lines (black lines). The size of each node is scaled by the actual power injected or consumed at the node from small (low absolute power) to large (high absolute power). The histogram above the map shows the net power distribution of $P_i$ for generators (green) and consumers (red). All data are taken from the ELMOD-DE data set~\cite{EGE16}.}
\end{figure}
In Figure~\ref{fig:powerGridGermany}, we provide a visualization of the German ultra-high voltage power grid. In order to determine the net power consumption/generation $P_i$ for each node in Fig~\ref{fig:powerGridGermany} depicted in the inset above the map, the individual power generation and consumption at each node are compared. We obtain the net power distribution $P_i=({P_\text{total}}/{C_\text{Total}})C_{\text{off},i} - P_{\text{off},i}$ where $P_\text{Total}=36 \,\mathrm{GW}$ and $C_\text{Total}\approx 88.343 \,\mathrm{GW}$ are the off-peak power consumption and generation of the whole power grid network, respectively, and $C_{\text{off},i}$ and $P_{\text{off},i}$ are the off-peak power consumption and generation for each individual node, respectively. Thus power balance $\sum_i P_i=0$ is guaranteed. For further details refer to~\cite{TAH19,MEN14}.

\begin{figure}[h!]
	\centering
	\includegraphics[width=0.8\columnwidth]{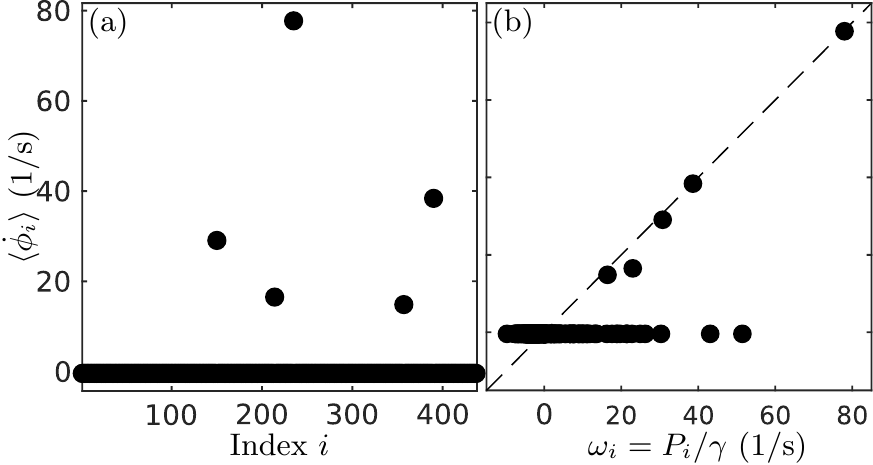}
	\caption{\label{fig:powerGridGermany_solitaryStates} (a) Mean phase velocity $\langle \dot{\phi}_i \rangle$ for each node in the German power grid network presented in Fig.~\ref{fig:powerGridGermany}. (b) Mean phase velocity  $\langle \dot{\phi}_i \rangle$ vs. natural phase velocity $\omega_i=P_i/\gamma$ for each node in the German ultra-high voltage power grid. The dashed line shows the relation $\langle \dot{\phi}_i\rangle=\omega_i$. Parameters in model~\eqref{eq:KwI_2order}: $M=I\omega_G$ with $I=40\times 10^3\,\mathrm{kg}\,\mathrm{m}^2$ and $\omega_G=2\pi 50\, \mathrm{Hz}$, $\gamma=Ma$ with $a =2\, \mathrm{Hz}$, $\sigma=800\, \mathrm{MW}$, $\alpha=0$, $t=600$, uniformly distributed random initial conditions $\phi\in(0,2\pi)$, $\dot{\phi}\in (-1,1)$.}
\end{figure}
In Figure~\ref{fig:powerGridGermany_solitaryStates}, we show a solitary state obtained by the simulation of model~\eqref{eq:KwI_2order} with the parameters as described above for $t=600$ and uniformly distributed random initial conditions $\phi\in(0,2\pi)$, $\dot{\phi}\in (-1,1)$. The temporal averages of the oscillators' phase velocities are obtained by neglecting the transient period $t\in[0,500)$. Solitary states are special cases of multifrequency cluster states where only single nodes have a different frequency compared to the large background cluster~\cite{BER20c}.  Figure~\ref{fig:powerGridGermany_solitaryStates} (a) shows such a solitary state where $5$ solitary nodes have a significantly different mean phase velocities than all the other oscillators from the large coherent cluster, which is synchronized at $\Omega_0\approx-0.407\,\mathrm{Hz}$. Similar results have been recently obtained in~\cite{TAH19,HEL20}. Remarkably, the mean phase velocities of the solitary nodes is very close to their natural frequency, see Fig.~\ref{fig:powerGridGermany_solitaryStates}(b). This means that the solitary states decouple on average from the mean field of their neighborhood, i.e., $\dot{\phi}_\text{Solitary}=\omega_\text{Solitary}
+\sum_j a_{ij} \chi_{ij}$ with temporal average $\langle \sum_j a_{ij} \chi_{ij} \rangle$ small compared to $\omega_\text{Solitary}$.

In order to shed light on further characteristics of the solitary states, we consider the power flows, i.e, the elements of the pseudo coupling matrix $\chi_{ij}$ introduced in~\eqref{eq:KwI_pseudo_kappa}.
\begin{figure}
	\centering
	\includegraphics[width=0.95\columnwidth]{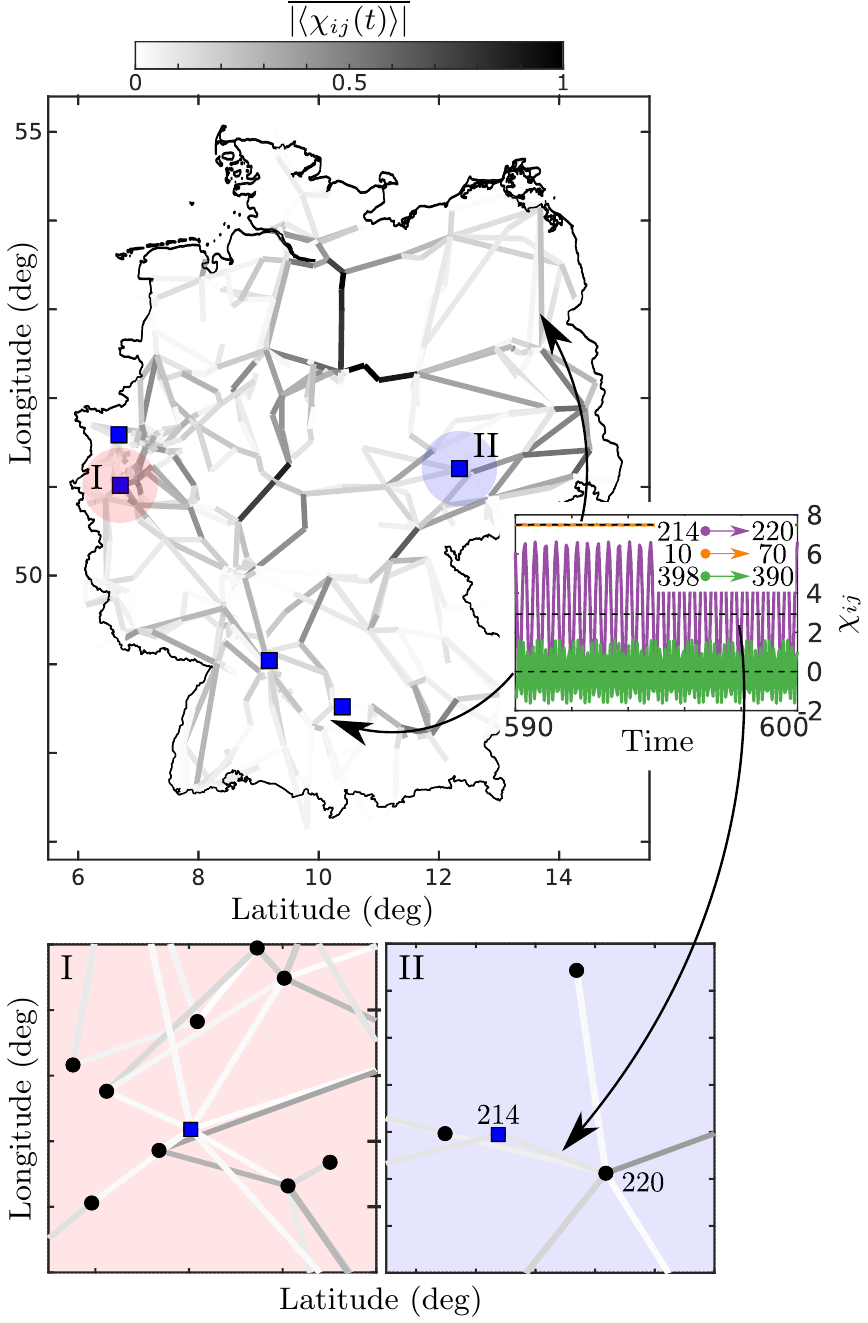}
	\caption{\label{fig:powerGridGermany_SolitaryPseudoCoup} Net power flow in the German ultra-high voltage power grid as given in Fig.~\ref{fig:powerGridGermany}. The solitary nodes presented in Fig.~\ref{fig:powerGridGermany_solitaryStates} are displayed as blue squares. For each transmission line, the grayscale shows the normalized average pseudo coupling weight $\overline{|\langle\chi_{ij}(t)\rangle|}$. The bar denotes the normalization for each value to the maximum for all $i,j=1,\dots,N$. The temporal evolution is evaluated over an averaging window of $100$ time units. The inset of the upper panel depicts the temporal evolution of three typical elements from the pseudo coupling matrix $\chi$, introduced in~\eqref{eq:KwI_pseudo_kappa}, with different mean power flow levels. Arrows point to the corresponding transmission lines. The dashed black lines show the values of the average pseudo coupling weights. Panels \rom{1} and \rom{2} provide blow-ups of the upper panel for the solitary nodes $i=235$ (pink shading) and $i=214$ (light blue shading), respectively. The black nodes in the blow-ups represent the consumers and generators of the power grid network. All parameters are as in Fig.~\ref{fig:powerGridGermany_solitaryStates}.
	}
\end{figure}
In Figure~\ref{fig:powerGridGermany_SolitaryPseudoCoup}, we provide an overview of the pseudo coupling matrix for the results obtained in the simulation of the German ultra-high voltage power grid, see also Fig.~\ref{fig:powerGridGermany_solitaryStates}. Note that we do not mark the nodes of the network in Fig.~\ref{fig:powerGridGermany_SolitaryPseudoCoup} (locations of the generators and consumers) to better visualize the characteristics of the pseudo coupling weights. We present the average coupling weights in Figure~\ref{fig:powerGridGermany_SolitaryPseudoCoup}. As we know from the discussion in Sec.~\ref{sec:dynRel}, the coupling weights correspond to the dynamical power flow of each transmission line. We further know that the average value of the power flow between a solitary node and a node from the coherent cluster is small but not necessarily zero. This is in fact supported by Fig.~\ref{fig:powerGridGermany_SolitaryPseudoCoup}, see also blow-ups I and II.

\begin{figure}
	\centering
	\includegraphics[width=0.95\columnwidth]{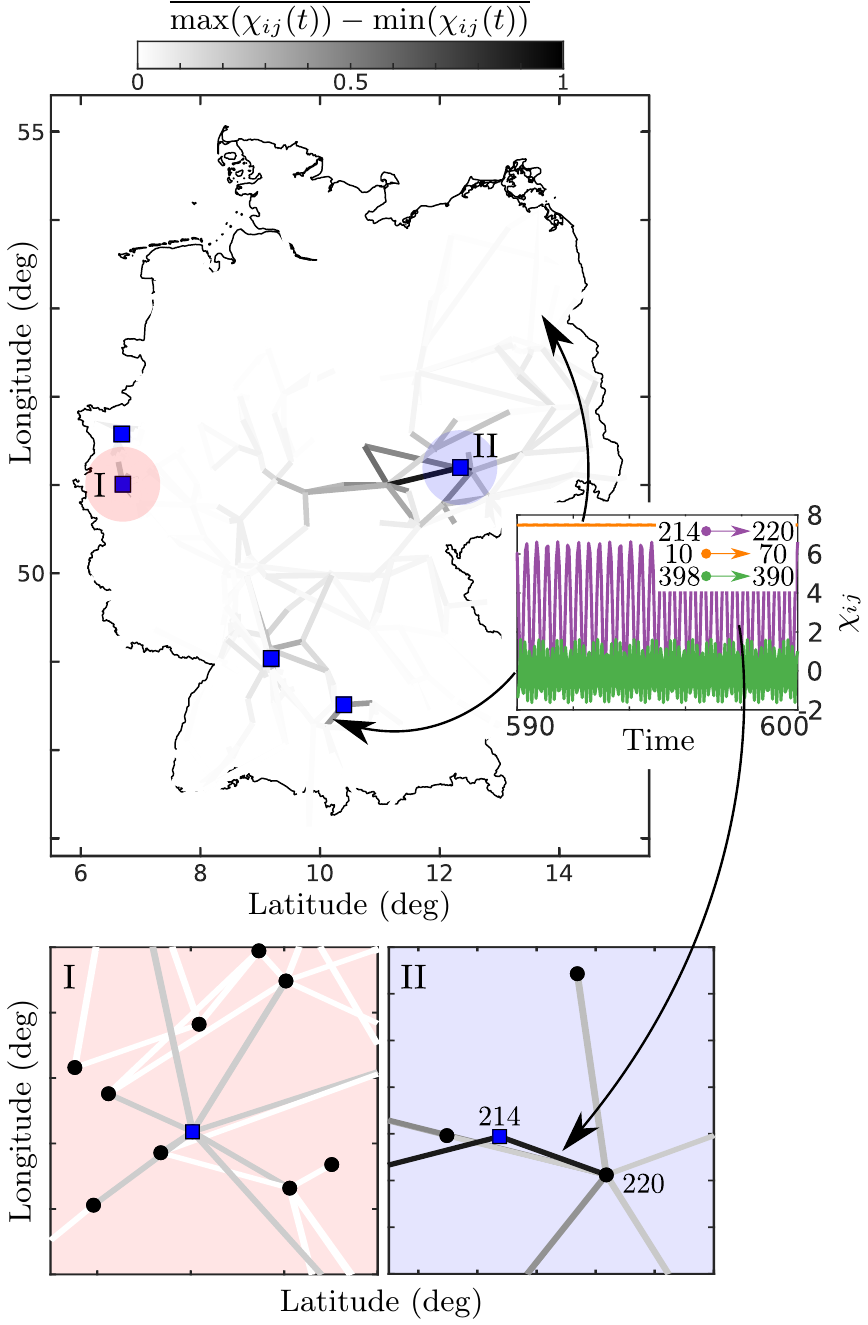}
	\caption{\label{fig:powerGridGermany_SolitaryPseudoCoupAmp} Variation of the net power flow in the German ultra-high voltage power grid. The solitary nodes presented in Fig.~\ref{fig:powerGridGermany_solitaryStates} are displayed as blue squares. For each transmission line, the grayscale shows the normalized amplitude of $\chi_{ij}(t)$, i.e., $\overline{\max(\chi_{ij}(t))-\min(\chi_{ij}(t))}$. %The bar denotes the normalization for each value to the maximum for all $i,j=1,\dots,N$. The temporal evolution is evaluated over an averaging window of $100$ time units. The inset of the upper panel depicts the temporal evolution of three elements of the pseudo coupling matrix $\chi$, introduced in~\eqref{eq:KwI_pseudo_kappa}, with different power flow variations. Arrows point to the corresponding transmission lines. Panels \rom{1} and \rom{2} provide blow-ups of the upper panel for the solitary nodes $i=235$ (pink shading) and $i=214$ (blue shading), respectively. The black nodes in the blow-ups represent the consumers and generators of the power grid network. 
The presentation is analogous to Fig.~\ref{fig:powerGridGermany_SolitaryPseudoCoup}. All parameters are as in Fig.~\ref{fig:powerGridGermany_solitaryStates}.}
\end{figure}

The temporal variations of the power flow are presented in Figure~\ref{fig:powerGridGermany_SolitaryPseudoCoupAmp}. Here, only a few lines show significant temporal variations. In particular, these lines are between solitary nodes and the coherent cluster. The blow-ups support the latter observation by showing the highest values of the temporal variation of the power flow for lines from and to the solitary nodes. Besides, Fig.~\ref{fig:powerGridGermany_SolitaryPseudoCoupAmp} shows how far into the network power flow fluctuations are spread in the presence of solitary states. It is visible that high power fluctuations exist even between nodes of the coherent cluster. These fluctuations would not be present if all oscillators were synchronized.

The insets in the upper panels of Figures~\ref{fig:powerGridGermany_SolitaryPseudoCoup} and~\ref{fig:powerGridGermany_SolitaryPseudoCoupAmp} depict the temporal evolution of three representative pseudo coupling weights. The three coupling weights vary periodically in time but with different amplitudes. For the coupling between two nodes of the coherent cluster ($10 \to 70$), the small variations stem from the small difference in their individual temporal dynamics which depends on their natural frequencies and the individual topological neighborhoods. In this realistic setup the dynamical network is very heterogeneous. In contrast to the case of two nodes of the coherent clusters, the couplings between solitary nodes and a node from the coherent cluster vary much more strongly periodically in time. To understand this observation, we derive an asymptotic approximation for the dynamics of the solitary states. Using an approach similar to~\cite{BER19}, the large power flow variations on transmission lines connecting solitary nodes can be explained. We apply a multiscale ansatz in $\epsilon=1/K\ll 1$ to a two-cluster state. By the two-cluster state, we model the interaction of a solitary node with the coherent cluster where $\phi^1$ represents the phase of the solitary node with natural frequency $\omega$ and $\phi^2$ represents the phase of the coherent cluster with natural frequency $\Omega_0$. The pseudo coupling weights between the two clusters are denoted by $\chi_{\mu\nu}$ ($\mu,\nu=1,2$, $\mu\ne\nu$). The ansatz reads $\phi^\mu=\Omega_\mu(\tau_0,\tau_1,\dots) + \epsilon(\phi^{(1,\mu)}) + \cdots$ and $\chi_{\mu\nu}=\chi_{\mu\nu}^{(0)}+\epsilon\chi_{\mu\nu}^{(1)}+ \cdots$ with $\tau_p=\epsilon^p t$, $p\in\mathbb{N}$. 

Omitting technical details, in the first order approximation in $\epsilon$ we obtain $\phi^1=\omega- ({K}/{\omega^2})\cos(\omega t)$ and $\phi^2=\Omega_0+ ({K}/{\omega^2})\cos(\omega t)$~\cite{BER20a}. Additional corrections to the oscillator frequencies appear in the third and higher orders of the expansion in $\epsilon$ and depend explicitly on $\omega$. The latter fact is consistent with the numerical observation in Fig.~\ref{fig:powerGridGermany_solitaryStates}(b) that solitary nodes with a lower natural frequency may differ more strongly from their own natural frequency than the solitary oscillators with a higher natural frequency.

As we have seen, the pseudo coupling approach allows for a description of the power flow for each line individually. It shows the emergence of large power flow fluctuations at the solitary nodes, and the spreading of those fluctuations over the power grid.

%--------------------------------------------------
% Conclusions
%--------------------------------------------------
\section{Conclusions}\label{sec:conclusions}
We have discussed the striking relation between phase oscillators with inertia, which are widely used for modeling power grids~\cite{ROH12,COR13,MOT13a,MEN14,WIT16,AUE17,MEH18,JAR18,SCH18c,SCH18i,TAH19,HEL20,KUE19,TOT20,MOL21} and adaptive networks of phase oscillators, which have ubiquitous applications in physical, biological, socioeconomic or neuronal systems. The introduction of the pseudo coupling matrix allows us to split the total input from all nodes into node $i$ into power flows. Thus the frequency deviation $\psi_i=\dot{\phi_i}-\omega_i$ in the phase oscillator model with inertia corresponds to the adaptively adjusted total input which an oscillator receives. This gives insight into the concept of phase oscillator models with inertia, which effectively takes into account the feedback loop of self-adjusted coupling with all other oscillators. Additionally, our novel theoretical framework allows for a generalization to swing equations with voltage dynamics~\cite{SCH14m}.

Our first example shows that the theory of building blocks developed for adaptively coupled phase oscillators can be transferred to explain the emergence of multicluster states in networks of coupled phase oscillators with inertia. These findings are of crucial importance for studying power grid models with respect to emergent multistability and dynamical effects that lead to desynchronization~\cite{PEC14,BAL19,ANV20}.

In fact, a properly functioning real-world power grid should be completely synchronized, i.e., clustering into different groups with different frequencies would be undesirable. However, multicluster states can still have practical relevance, since they influence the destabilization of the synchronous state. Thus, it is important to study when they occur, in order to be able to take control measures to prevent them. For instance, recent works~\cite{TAH19,HEL20} have shown that the solitary states, which are a subclass of multicluster states, arise naturally in the desynchronization transition of real-world power grid networks (German and Scandinavian power grid), and that this knowledge is essential for an efficient power grid control. For the German power grid, we provide an additional example and show analytically how the techniques developed for adaptive networks are used to characterize the emergent solitary states. We have shown that the concept of pseudo coupling weights is powerful tool to analyze the dynamical spreading of power flow fluctuations. Therefore, we believe that this concept can be used in the future to design novel detection and control approaches for modern power grid networks. 

%\section*{Acknowledgment}

%This work was supported by the German Research Foundation DFG, Project Nos. 411803875 and 440145547. 

%\bibliography{ref}

\begin{thebibliography}{10}
	\expandafter\ifx\csname url\endcsname\relax
	\def\url#1{{\tt #1}}\fi
	\expandafter\ifx\csname urlprefix\endcsname\relax\def\urlprefix{URL }\fi
	
	\bibitem{SAU98a}
	P.~W. Sauer and M.~A. Pai: {\em Power system dynamics and stability\/}, vol.
	101 (Upper Saddle River, NJ: Prentice hall, 1998).
	
	\bibitem{FIL08a}
	G.~Filatrella, A.~H. Nielsen, and N.~F. Pedersen: {\em Analysis of a power grid
		using a {K}uramoto-like model\/}, Eur. Phys. J. B {\bf 61}, 485 (2008).
	
	\bibitem{SCH16o}
	J.~Schiffer, D.~Zonetti, R.~Ortega, and A.~M. Stankovic: {\em A survey on
		modeling of microgrids-from fundamental physics to phasors and voltage
		sources\/}, Automatica {\bf 74}, 135 (2016).
	
	\bibitem{DOE13}
	F.~D\"orfler, M.~Chertkov, and F.~Bullo: {\em Synchronization in complex
		oscillator networks and smart grids\/}, Proc. Natl. Acad. Sci. U.S.A. {\bf
		110}, 2005 (2013).
	
	\bibitem{NIS15}
	T.~Nishikawa and A.~E. Motter: {\em Comparative analysis of existing models for
		power-grid synchronization\/}, New J. Phys. {\bf 17}, 015012 (2015).
	
	\bibitem{AUE16}
	S.~Auer, K.~Kleis, P.~Schultz, J.~Kurths, and F.~Hellmann: {\em The impact of
		model detail on power grid resilience measures\/}, Eur. Phys. J. Spec. Top.
	{\bf 225}, 609 (2016).
	
	\bibitem{DOE14}
	F.~D\"orfler and F.~Bullo: {\em Synchronization in complex networks of phase
		oscillators: A survey\/}, Automatica {\bf 50}, 1539 (2014).
	
	\bibitem{ROD16}
	F.~A. Rodrigues, T.~K. D.~M. Peron, P.~Ji, and J.~Kurths: {\em The {Kuramoto}
		model in complex networks\/}, Phys. Rep. {\bf 610}, 1 (2016).
	
	\bibitem{ROH12}
	M.~Rohden, A.~Sorge, M.~Timme, and D.~Witthaut: {\em Self-organized
		synchronization in decentralized power grids\/}, Phys. Rev. Lett. {\bf 109},
	064101 (2012).
	
	\bibitem{COR13}
	S.~P. Cornelius, W.~L. Kath, and A.~E. Motter: {\em Realistic control of
		network dynamics\/}, Nat. Commun. {\bf 4}, 1942 (2013).
	
	\bibitem{MOT13a}
	A.~E. Motter, S.~A. Myers, M.~Anghel, and T.~Nishikawa: {\em Spontaneous
		synchrony in power-grid networks\/}, Nat. Phys. {\bf 9}, 191 (2013).
	
	\bibitem{MEN14}
	P.~J. Menck, J.~Heitzig, J.~Kurths, and H.~J. Schellnhuber: {\em How dead ends
		undermine power grid stability\/}, Nat. Commun. {\bf 5}, 3969 (2014).
	
	\bibitem{WIT16}
	D.~Witthaut, M.~Rohden, X.~Zhang, S.~Hallerberg, and M.~Timme: {\em Critical
		links and nonlocal rerouting in complex supply networks\/}, Phys. Rev. Lett.
	{\bf 116}, 138701 (2016).
	
	\bibitem{AUE17}
	S.~Auer, F.~Hellmann, M.~Krause, and J.~Kurths: {\em Stability of synchrony
		against local intermittent fluctuations in tree-like power grids\/}, Chaos
	{\bf 27}, 127003 (2017).
	
	\bibitem{MEH18}
	V.~Mehrmann, R.~Morandin, S.~Olmi, and E.~Sch{\"o}ll: {\em Qualitative
		stability and synchronicity analysis of power network models in
		port-{H}amiltonian form\/}, Chaos {\bf 28}, 101102 (2018).
	
	\bibitem{JAR18}
	P.~Jaros, S.~Brezetsky, R.~Levchenko, D.~Dudkowski, T.~Kapitaniak, and
	Y.~Maistrenko: {\em Solitary states for coupled oscillators with inertia\/},
	Chaos {\bf 28}, 011103 (2018).
	
	\bibitem{SCH18c}
	B.~Sch\"afer, C.~Beck, K.~Aihara, D.~Witthaut, and M.~Timme: {\em
		Non-{G}aussian power grid frequency fluctuations characterized by
		{L}evy-stable laws and superstatistics\/}, Nature Energy {\bf 3}, 119 (2018).
	
	\bibitem{SCH18i}
	B.~Sch\"afer, D.~Witthaut, M.~Timme, and V.~Latora: {\em Dynamically induced
		cascading failures in power grids\/}, Nat. Commun. {\bf 9}, 1975 (2018).
	
	\bibitem{TAH19}
	H.~Taher, S.~Olmi, and E.~Sch{\"o}ll: {\em Enhancing power grid synchronization
		and stability through time delayed feedback control\/}, Phys. Rev. E {\bf
		100}, 062306 (2019).
	
	\bibitem{HEL20}
	F.~Hellmann, P.~Schultz, P.~Jaros, R.~Levchenko, T.~Kapitaniak, J.~Kurths, and
	Y.~Maistrenko: {\em Network-induced multistability through lossy coupling and
		exotic solitary states\/}, Nat. Commun. {\bf 11}, 592 (2020).
	
	\bibitem{KUE19}
	C.~Kuehn and S.~Throm: {\em Power network dynamics on graphons\/}, SIAM J.
	Appl. Dyn. Syst. {\bf 79}, 1271 (2019).
	
	\bibitem{TOT20}
	C.~H. Totz, S.~Olmi, and E.~Sch{\"o}ll: {\em Control of synchronization in
		two-layer power grids\/}, Phys. Rev. E {\bf 102}, 022311 (2020).
	
	\bibitem{MOL21}
	F.~Molnar, T.~Nishikawa, and A.~E. Motter: {\em Asymmetry underlies stability
		in power grids\/}, Nat. Commun. {\bf 12}, 1457 (2021).
	
	\bibitem{BER20c}
	R.~Berner, A.~Polanska, E.~Sch{\"o}ll, and S.~Yanchuk: {\em Solitary states in
		adaptive nonlocal oscillator networks\/}, Eur. Phys. J. Spec. Top. {\bf 229},
	2183 (2020).
	
	\bibitem{BEL16a}
	I.~V. Belykh, B.~N. Brister, and V.~N. Belykh: {\em Bistability of patterns of
		synchrony in {Kuramoto} oscillators with inertia\/}, Chaos {\bf 26}, 094822
	(2016).
	
	\bibitem{BER19}
	R.~Berner, E.~Sch{\"o}ll, and S.~Yanchuk: {\em Multiclusters in networks of
		adaptively coupled phase oscillators\/}, SIAM J. Appl. Dyn. Syst. {\bf 18},
	2227 (2019).
	
	\bibitem{BER19a}
	R.~Berner, J.~Fialkowski, D.~V. Kasatkin, V.~I. Nekorkin, S.~Yanchuk, and
	E.~Sch{\"o}ll: {\em Hierarchical frequency clusters in adaptive networks of
		phase oscillators\/}, Chaos {\bf 29}, 103134 (2019).
	
	\bibitem{TUM19}
	L.~Tumash, S.~Olmi, and E.~Sch{\"o}ll: {\em {S}tability and control of power
		grids with diluted network topology\/}, Chaos {\bf 29}, 123105 (2019).
	
	\bibitem{OLM15a}
	S.~Olmi: {\em Chimera states in coupled {Kuramoto} oscillators with inertia\/},
	Chaos {\bf 25}, 123125 (2015).
	
	\bibitem{KAS17}
	D.~V. Kasatkin, S.~Yanchuk, E.~Sch{\"o}ll, and V.~I. Nekorkin: {\em
		{S}elf-organized emergence of multi-layer structure and chimera states in
		dynamical networks with adaptive couplings\/}, Phys. Rev. E {\bf 96}, 062211
	(2017).
	
	\bibitem{OLM14a}
	S.~Olmi, A.~Navas, S.~Boccaletti, and A.~Torcini: {\em Hysteretic transitions
		in the {Kuramoto} model with inertia\/}, Phys. Rev. E {\bf 90}, 042905
	(2014).
	
	\bibitem{ZHA15a}
	X.~Zhang, S.~Boccaletti, S.~Guan, and Z.~Liu: {\em Explosive synchronization in
		adaptive and multilayer networks\/}, Phys. Rev. Lett. {\bf 114}, 038701
	(2015).
	
	\bibitem{BAR16a}
	J.~Barr{\'e} and D.~M{\'e}tivier: {\em Bifurcations and singularities for
		coupled oscillators with inertia and frustration\/}, Phys. Rev. Lett. {\bf
		117}, 214102 (2016).
	
	\bibitem{TUM18}
	L.~Tumash, S.~Olmi, and E.~Sch{\"o}ll: {\em {E}ffect of disorder and noise in
		shaping the dynamics of power grids\/}, Europhys. Lett. {\bf 123}, 20001
	(2018).
	
	\bibitem{YAN17a}
	Y.~Yang and A.~E. Motter: {\em Cascading {F}ailures as {C}ontinuous
		{P}hase-{S}pace {T}ransitions\/}, Phys. Rev. Lett. {\bf 199}, 248302 (2017).
	
	\bibitem{SCH14m}
	K.~Schmietendorf, J.~Peinke, R.~Friedrich, and O.~Kamps: {\em Self-organized
		synchronization and voltage stability in networks of synchronous machines\/},
	Eur. Phys. J. Spec. Top. {\bf 223}, 2577 (2014).
	
	\bibitem{CIS20}
	M.~{Ciszak}, F.~Marino, A.~Torcini, and S.~Olmi: {\em Emergent excitability in
		populations of nonexcitable units\/}, Phys. Rev. E {\bf 102}, 050201(R)
	(2020).
	
	\bibitem{BER20a}
	R.~Berner, S.~Yanchuk, and E.~Sch{\"o}ll: {\em What adaptive neuronal networks
		teach us about power grids\/}, Phys. Rev. E {\bf in print} (2021),
	arXiv:2006.06353.
	
	\bibitem{SAK86}
	H.~Sakaguchi and Y.~Kuramoto: {\em A soluble active rotater model showing phase
		transitions via mutual entertainment\/}, Prog. Theor. Phys {\bf 76}, 576
	(1986).
	
	\bibitem{BER20b}
	R.~Berner, S.~Vock, E.~Sch{\"o}ll, and S.~Yanchuk: {\em Desynchronization
		transitions in adaptive networks\/}, Phys. Rev. Lett. {\bf 126}, 028301
	(2021).
	
	\bibitem{KUR84}
	Y.~Kuramoto: {\em Chemical Oscillations, Waves and Turbulence\/}
	(Springer-Verlag, Berlin, 1984).
	
	\bibitem{ABR96}
	W.~C. Abraham and M.~F. Bear: {\em Metaplasticity: the plasticity of synaptic
		plasticity\/}, Trends Neurosci. {\bf 19}, 126 (1996).
	
	\bibitem{ABR08a}
	W.~C. Abraham: {\em Metaplasticity: tuning synapses and networks for
		plasticity\/}, Nat. Rev. Neurosci. {\bf 9}, 387 (2008).
	
	\bibitem{JOH18}
	R.~A. John, F.~Liu, N.~A. Chien, M.~R. Kulkarni, C.~Zhu, Q.~D. Fu, A.~Basu,
	Z.~Liu, and N.~Mathews: {\em Synergistic gating of electro-iono-photoactive
		2d chalcogenide neuristors: Coexistence of hebbian and homeostatic synaptic
		metaplasticity\/}, Adv. Mater. {\bf 30}, 1800220 (2018).
	
	\bibitem{EGE16}
	J.~Egerer: {\em {Open Source Electricity Model for Germany (ELMOD-DE)}\/},
	Tech. rep., Deutsches {I}nstitut f\"{u}r {W}irtschaftsforschung {(DIW)}
	(2016).
	
	\bibitem{PEC14}
	L.~M. Pecora, F.~Sorrentino, A.~M. Hagerstrom, T.~E. Murphy, and R.~Roy: {\em
		Symmetries, cluster synchronization, and isolated desynchronization in
		complex networks\/}, Nat. Commun. {\bf 5}, 4079 (2014).
	
	\bibitem{BAL19}
	C.~Balestra, F.~Kaiser, D.~Manik, and D.~Witthaut: {\em Multistability in lossy
		power grids and oscillator networks\/}, Chaos {\bf 29}, 123119 (2019).
	
	\bibitem{ANV20}
	M.~Anvari, F.~Hellmann, and X.~Zhang: {\em Introduction to focus issue:
		Dynamics of modern power grids\/}, Chaos {\bf 30}, 063140 (2020).
	
\end{thebibliography}
%\bibliographystyle{prwithtitle}

\end{document}